\documentclass{elsart}
\usepackage{graphicx,amssymb}
\usepackage[square,comma]{natbib}
\journal{Physica C}
\begin{document}
\begin{frontmatter}

\title{High--$T_c$ RF SQUIDs\\ with Large Inductance of Quantization Loop}
\author[]{A.V. Sermyagin}
\ead{sermyagin@jinr.ru}
\address{Institute in Physical--Technical Problems,
Dubna, Russia}
\begin{abstract}
Experimental data on signal and noise characteristics of
high--$T_c$ RF SQUIDs with large inductance of quantization loop
are presented. The SQUIDs were produced by a thick--film
HTS--technique of painting on the Y$_2$BaCuO$_5$ substrate.  For
the first time, a steady quantum interference was observed in RF
SQUID with the inductance as large as $L_S=6.6$ nH, that is, 60
times higher than the fluctuation inductance $L_F\simeq10^{-10}H$
at the liquid nitrogen temperature $T=77$ K. A new method is
offered to evaluate the sensitivity of RF SQUID with optimum
inductance of the quantization loop.
\end{abstract}
\end{frontmatter}
\maketitle
\section*{Introduction}
\begin{figure}[t]
\begin{center}
\scalebox{1.0}{\includegraphics[1cm,27cm][15cm,0cm]{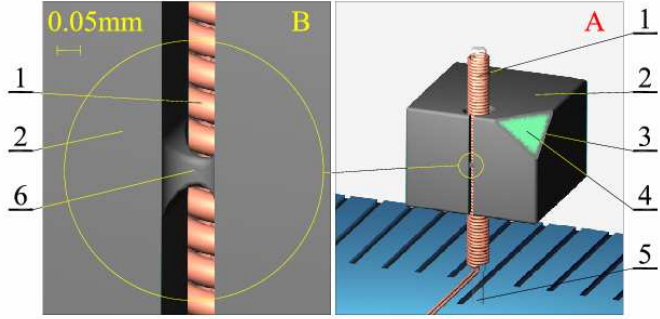}}
\vspace{4.5cm}\end{center} \caption{{\it TF RF SQUID, made by the
technique of painting, A, and a magnified view of the SQUID
bridge, B.} 1 -- {\it the coupling coil;} 2 -- {\it the SQUID
sensor body;} {\it on the angle truncated:} 3 -- {\it the layer
of} YBa$_2$Cu$_3$O$_{7-\delta}$, 4 -- {\it the substrate of}
Y$_2$BaCuO$_5$; 5 -- {\it the coupling coil sheave;} 6 -- {\it the
SQUID bridge.}}\label{fig1.eps}
\end{figure}

At the beginning of the 90th, at the Institute in
Physical--Technical Problems, Dubna, an original way of
preparation high--temperature superconducting (HTS) YBa$_2$Cu$_
3$O$_{7-\delta}$ thick films (TFs) by a quasi binary technique
\cite{Vuong_1993}, was developed. The superconducting layer of
YBa$_2$Cu$_3$O$_{7-\delta}$ was formed during the reaction of
synthesis between the chemically active Y$_2$BaCuO$_5$ substrate
and the layer of ``paint'' prepared from a mixture of BaCuO$_2$
and CuO, which was applied to the substrate.\\\indent A
modification of this technique by means of the melting--growing
(MG) technology~\cite{Murakami_1992} allowed one to obtain the TFs
with microscopic inclusions of the Y$_2$BaCuO$_5$--phase, which
play the role of additional pinning centers in the
YBa$_2$Cu$_3$O$_{7-\delta}$--matrix~\cite{Vuong_1994}.\\\indent
Temperature and field dependency study of the AC susceptibility of
these MG TF structures shows that pinning force is increased
several times as compared with TF reference samples (without MG)
which superconductive characteristics are similar to those of the
samples synthesized by the conventional ceramic
technique~\cite{Vuong_1994}. The high density (minimum porosity)
is an important feature of the MG TFs, and determines higher
stability of such structures against a degradation process. Next,
one--hole RF SQUIDs (see Fig.~\ref{fig1.eps}) were built on the
basis of the MG TFs with the help of a developed method of making
SQUID bridges (by a degradation in hot water vapour), and their
main characteristics were
investigated~\cite{Vasiliev1995}.\\\indent At the operating
(pumping) frequency 20\,MHz, the best performance of SQUID
parameters with $L_S=29{.}7$\,pH was obtained:  the spectral noise
density (s.\,n.\,d.) $S_\varphi\sim 3\cdot
10^{-4}\Phi_{0}$/Hz$^{1/2}$, where $\Phi_0$ is the flux quanta,
the energy resolution $\varepsilon\sim 6{.}5\cdot 10^{-27}$\,J/Hz,
the cutoff frequency of the excess $1/f$ noise was less than
$1$\,Hz, and the field sensitivity $S_F\sim 3\cdot
10^{-13}$\,T/Hz$^{1/2}$.
These studies did not aim to develope an RF SQUID with the large
inductance; such SQUIDs were a by--product of works, i.\,e. a
consequence of painting defects that led to the disorganized
geometry of a quantization loop but not to degradation of
superconductive properties as well.
\begin{figure}[t]
\begin{center}
\scalebox{0.45}{\includegraphics[15cm,26.5cm][14cm,0cm]{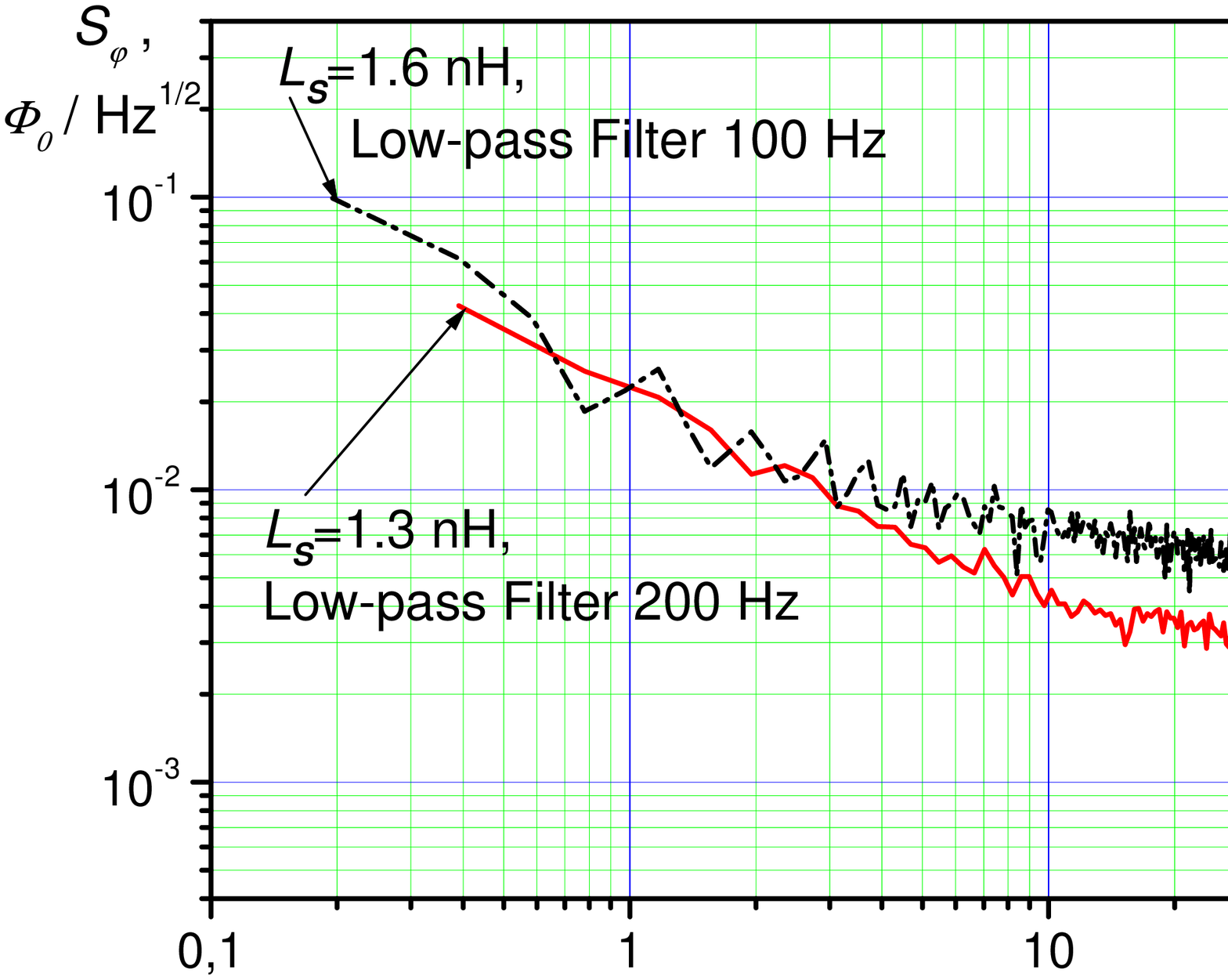}}
\vspace{7cm}\end{center}\caption{\it S.\,n.\,d. of SQUIDs with
$L_S=1{.}3$\,nH and $L_S=1{.}6$\,nH.} \label{Spectr}
\end{figure}
It is clear now that it is precisely these good superconductive
properties of the high--$T_c$ TFs synthesized in this way, and
their high stability to degradation, that allowed us to obtain the
data for s.\,n.\,d. of several large inductance SQUIDs with
$L_S\gg L_F$ (see Fig.~\ref{Spectr}).\\\indent In these SQUIDs,
the popular pattern of the steady state quantum interference, as
it is shown in Fig.~\ref{TipicalVF}, was observed.\\
\begin{figure}[b]
\begin{center}
\scalebox{0.5}{\includegraphics[13cm,17.5cm][10.5cm,0cm]{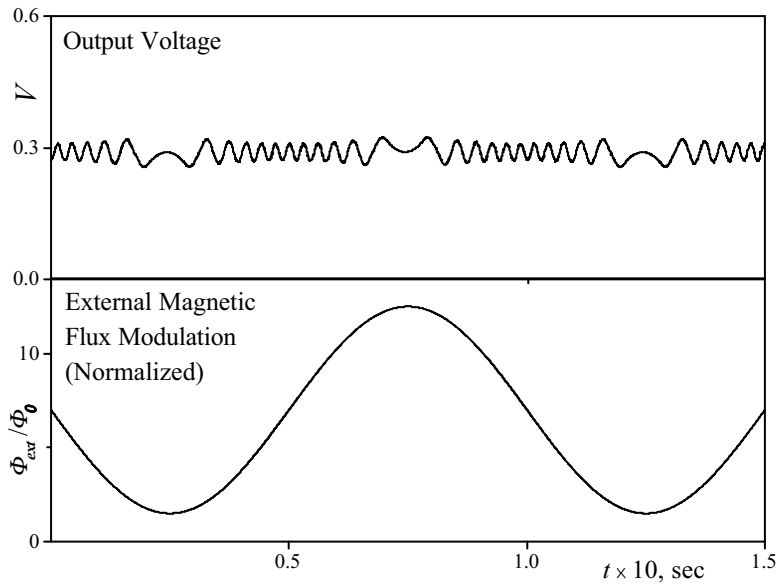}}
\vspace{7.5cm}\end{center}\caption{\it Typical volt--to--flux
characteristic of RF SQUID having large inductance $L_S$ of
quantization loop, obtained with sine--wave modulation of external
magnetic flux (digital modeling).} \label{TipicalVF}
\end{figure}
\indent{\it Large inductance SQUIDs.} Usually, SQUIDs with obvious
defects of uniformity of TF, such as the unpainted sites on their
surface, manifest high magnitude of quantization loop inductance,
which is much higher than the calculated magnitude of geometrical
SQUID--hole inductance.\\\indent In the case of biconnected
samples, similar phenomenon is considered commonly as a
manifestation of the percolation nature of supercurrent in
ceramics with poor content of superconducting
phase~\cite{Verkin}{.}\\\indent In our case, this fact can be
interpreted as follows. The one--hole SQUID is not a
one--inductance device; in this regard, it is similar to a
two--hole Zimmerman's SQUID~\cite{Zimmerman1970}{,} but, unlike
that, the one--hole RF SQUID should be considered as an asymmetric
one, including two different loops of inductance.\\\indent The
main, or the master, quantization loop is $L_S$ loop (see
Fig.~\ref{fig4.eps}) located on the {\it internal} surface of the
SQUID hole. When the SQUID bridge occurs in a normal resistive
state under the action of pumping, the inductance $L_S$ joins in
series the loop $L_C$ located on the {\it external} surface of
SQUID (see Fig.~\ref{fig4.eps}). If there are no defects, then
$L_C\sim L_S$ (see Fig.~\ref{fig4.eps}~A). One can see that $L_C$
and $L_S$ are connected in a subtractive polarity mode, then such
a contour operates as a differential flux transformer having a
small effective inductance and a low field sensitivity, as was
observed in~\cite{Vasiliev1995}. The defects of painting as is
shown in Fig.~\ref{fig4.eps}~B, may yield $L_C\gg L_S$; then, the
effective inductance of RF SQUID is determined by the external
loop inductance $L_C$. The kinetic inductance, which influence in
the case of thin--film is considered essential~\cite {VVS}{,} may
give some contribution to the impedance of the quantization loop;
in general, this aspect needs a separate research (see,
also~\cite{Koelle_Chesca,Sermyagin_2003}).\\\indent
\begin{figure}[t]
\begin{center}
\scalebox{1.0}{\includegraphics[9cm,16cm][12cm,0cm]{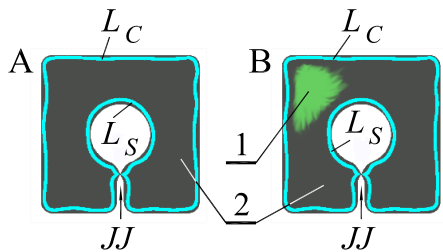}}
\vspace{2.5cm}\end{center}\caption{{\it Topology of the
quantization loop of the one--hole RF SQUID.} A -- {\it A
defect--free high--$T_c$ sample.} B -- {\it A sample containing
non--superconductive superficial } Y$_2$BaCuO$_5$ {\it
inclusions}, 1; 2 -- YBa$_2$Cu$_3$O$_{7-\delta}$ {\it
superconducting layer}; $JJ$ -- {\it SQUID bridge (Josephson
junction).} {\it For explanation of }$L_C$ {\it and} $L_S$ {\it
see the text.}} \label{fig4.eps}
\end{figure}
Hereinafter $L_S$ means the RF SQUID efficient quantization loop
inductance, which can be evaluated experimentally.\\\indent For an
experimental determination of RF SQUID parameters, techniques
\cite{Polushkin1989_JINR_Rapid, Polushkin1990} were employed. The
techniques allow one to find the SQUID inductance $L_S$ and its
coupling coefficient $k$ between the interferometer loop and the
tank circuit, with the accuracy approximately of 20$\%$. A
critical current $I_c$ of weak link is calculated directly from
the value of the hysteresis parameter $\beta=2\pi(I_cL_S/\Phi_0)$
defined from the measuring of the amplitude and displacement of
the flux--to--voltage
characteristic~\cite{Likharev1985}{.}\\\indent In this paper,
experimental data for RF SQUIDs with large and very large
inductances $L_S\gg L_F$, considerably exceeding the known
published values~\cite{Polushkin1989, Vasiliev1990, Il'ichev1997,
Chesca_1998}{,} are presented. A derivation of a new
flux--resolution formula for $S_\varphi$ depending on $L_S$, which
seems to be correct over the broad range $L_S\gg L_F$, is given.
It is shown, by applying the relationship for $S_\varphi(L_S)$,
that experimental data for s.\,n.\,d. of a large inductance RF
SQUID allow one to estimate a sensitivity of the RF SQUID with the
loop inductance $L_S=L_F$; i.\,e., to actually determine  a
sensitivity of the RF SQUID made of a given HTS in the case of
optimal quantization loop inductance.
\section{Fluctuations in the Interferometer}
An important inherent feature of HTS SQUIDs working at 77 K, is a
high level of fluctuations. If thermal fluctuations dominate in a
SQUID, a method of the Fokker--Plank equation should be
applied~\cite{Likharev1985, Polushkin1989, Chesca_1998}.\\\indent
The Fokker--Plank equation is written for the probability density
$\sigma(t,\varphi,v)$ to find a system at a time $t$, phase
$\varphi$, and volume $v$. Knowing $\sigma(t,\varphi,v)$, it is
possible to find the average of a function $F(\varphi,v)$ over the
statistical ensemble by the usual way
\begin{equation}\label{MeanValue}
  \langle F(t)\rangle=\int\limits_{-\infty}^\infty dv\int\limits_{-\infty}^\infty
  d\varphi\sigma(t,\varphi,v)F(\varphi,v).
\end{equation}
In a non--hysteretic ($\beta < 1$) SQUID, only the steady state
corresponding to the minimum of the Hibbs function $G$ exists,
i.\,e., the system energy
\begin{eqnarray}
\label{G}
G=K+U\phantom{\,\,\,=Q^2/2C+E_c[(1-\cos{\varphi})+(\varphi-\varphi_{ext})^2/2\beta]}\nonumber\\
 =Q^2/2C+E_c[(1-\cos{\varphi})+(\varphi-\varphi_{ext})^2/2\beta],\phantom{=K+U}
\end{eqnarray}
where $K$ and $U$ are the kinetic and the potential energy of the
SQUID, $Q$ and $C$ are the charge and the capacity of the
Josephson junction, accordingly, $E_c=\hbar I_c/2e$ is the
Josephson coupling energy of the junction,
$\varphi=2\pi\Phi/\Phi_0$, $\Phi$ is the magnetic flux coupled to
the quantization loop by the relationship
$\beta\sin{\varphi}=\varphi_{ext}-\varphi$,
$\varphi_{ext}=2\pi\Phi_{ext}/\Phi_0$, and $\Phi_{ext}$ is the
external magnetic flux.\\\indent A small noise yields, therefore,
small phase fluctuations within this state. The probability
density of finding a system in a phase space point is proportional
to the Bolzmann distribution
\begin{equation}\label{Bolzman}
  \sigma\sim\exp(-G/k_BT).
\end{equation}
If $ \beta\ll 1 $, it is enough to hold the last term in the
potential energy expression (\ref {G}), and for the dispersion of
flux fluctuations one receives
\begin{equation}\label{Noise}
\langle{\widetilde{\Phi}}^2\rangle\approx(\Phi_0/2\pi)^2\langle{\widetilde{\varphi}}^2\rangle\approx
k_BTL_S.
\end{equation}
The expression (\ref{Noise}) permits one to estimate the
interference degradation in the case of $L_S >
L_F$~\cite{Likharev1985}{,}
\begin{equation}\label{L>L_F}
\langle U
\rangle\approx-E_c\cos{\langle\varphi\rangle}\exp(-L_S/2L_F),
\end{equation}
where $L_F\equiv(\Phi_0/2\pi)^2(k_BT)^{-1}$; at $T=4$\,K,
$L_F=2\cdot 10^{-9}$\,H.\\\indent The relationship (\ref{L>L_F})
is valid for a nonhysteretic mode the criterion of which is the
validity of the inequality for the hysteretic SQUID parameter,
$\beta <1$, by the former theories which have been offered prior
to the discovery of HTS.\\\indent However as early as in 1975, in
the theoretical analysis of an RF SQUID it was
noticed~\cite{Khlus_Kulik}{,} that in the case of large
fluctuations, i.\,e., when the Josephson coupling energy $E_c$ of
junction is less than the average thermal energy
($\Gamma=(k_BT/E_c)>1$), the SQUID is nonhysteretic.
B.\,Chesca~\cite{Chesca_1998} and
Ya.\,Greenberg~\cite{Greenberg_1999} have come  to a similar
conclusion developing an RF SQUID theory applicable to the case of
high fluctuations and which conclusions are supported by the known
experimental data. In particular, according to the
Chesca--Greenberg theory, an RF SQUID is non hysteretic under
conditions of high fluctuations, up to the hysteresis parameter
value of $\beta\leq3$. Thus, there is a basis for applying a
simple form (\ref{L>L_F}) of the interference supression
dependence on inductance to estimate a noise of the SQUID with
high inductance $L_S$.\\\indent The components of a net noise of
an interferometer that determine its sensitivity, are as
follows~\cite{Rogachevskiy1993,Koelle}:
\begin{itemize}
\item self--noise, or internal noise of the sensor (SQUID) of the interferometer;
\item the noise of the tank circuit coupled inductively to the SQUID;
\item the noise of a feeder connecting the tank circuit to the preamplifier;
if the preamplifier is placed outside the dewar, the ends of the
feeder have a temperature difference causing the noise.
\item Noises of the first amplifier stage and of a resistor in the feedback
loop. They can be reduced sufficiently by cooling the first stage.
It is profitable, from the view point of increasing sensitivity,
in the case of low--$T_c$ SQUIDs having  the own sensor noise much
lower than the rest noise sources. For high--$T_c$ SQUIDs
operating at liquid nitrogen temperature, the sensor noise is not
much less than the noise of electronics.
\item The external noise. To estimate the interferometer sensitivity, it should be minimized.
The best way to minimize external influences is to apply a
superconducting shield the SQUID sensor is placed into. For this
purpose, the high--$T_c$ shield should have a low level of the
self--noise. This condition is met when the high--$T_c$ shield has
a high enough value of the penetrating
field~\cite{Polushkin_Buev}. Nevertheless, employing an ordinary
ferromagnetic shielding provides a relevant solution.
\end{itemize}
\par The internal noise of the SQUID sensor is defined by Josephson
junction properties, and by magnetic flux fluctuations having
dispersion (\ref{Noise}). Self--noise of HTS--based Josephson
junction can be low enough; the higher the critical current
density, the lower the self--noise. In YBCO systems, it is
possible to achieve high values of the pinning
force~\cite{Vuong_1994, Vasiliev1995, Koelle}{,} which guarantees
high values of critical current density in the $ab$--plane of a
YBCO crystal cell, and an acceptably low level of excess $1/f$
noise as well. A theoretical estimate of the internal noise level
and the maximum inductance of the RF SQUID quantization loop can
be carried out in a simple way~\cite{Polushkin1989_JINR_Rapid}.
For an RF SQUID with the inductance $L_S=10^{-10}$\,H the estimate
of s.\,n.\,d. at the temperature 77\,K gives
\begin{equation}\label{25}
S_\varphi=10^{-5}\mbox{\,}\Phi_0/\mbox{Hz}^{1/2}.
\end{equation}
According to~\cite{Polushkin1989_JINR_Rapid}{,} quantum
interference in the RF SQUID is hardly observed if the value of
the RF SQUID threshold inductance $L_{max}$ is exceeded
\begin{equation}\label{22}
L_S>L_{max}= \Phi^2_0/k_BT =4\cdot10^{-9}\mbox{\,H.}
\end{equation}
At that time (1989), these conclusions were not experimentally
proven probably due to poor quality of HTS ceramics, which did not
allow simple handling of SQUIDs. At present, the sensitivity level
(\ref {25}) is successfully achieved in high--$T_c$ RF SQUIDs
produced on the basis of modern thin film
technologies~\cite{Zhang}. In its turn, studying of the SQUIDs
made by using the thick--film technique, have a confirmation and
specification of these estimates as well. Thus, according to the
estimate (\ref {22}), for the high--$T_c$ SQUIDs operated at
$77$\,K, a wide enough inductance range including even
``He--values'' up to several nH is available. According to the
estimate (\ref {25}), in the case of the optimum quantization loop
inductance, the interferometer sensor self--noise is appreciably
below the tank circuit noise and the noise of the ``warm''
preamplifier, as well as in the case of the low--$T_c$ RF SQUIDs.
As evident from what follows, the estimates of the maximum
sensitivity of the RF SQUIDs with the optimum quantization loop
inductance, carried out on the basis of the experimental
s.\,n.\,d. data for large inductance SQUIDs, agree well with (\ref
{25}).
\section{Dual SQUID}
Let us introduce a concept of a SQUID which is dual to the RF
SQUID with a large quantization loop inductance. A {\it dual}
SQUID is a SQUID with the same weak link properties as a SQUID
with a high inductance value, but with the optimum quantization
loop inductance value equal to $L_F$. In a formal way, it is
possible to obtain a dual SQUID from a large inductance RF SQUID
by deforming the quantization loop to reduce its inductance down
to $L_F$, preserving the weak link parameters. Since the noise of
the SQUID sensor with the large inductance $L_S\gg L_F$ is
determined, in accordance with (\ref{L>L_F}), by the ``Debye''
factor $\exp{(L_S/2L_F)}$ only, it is possible to carry out a
simple theoretical estimate of the sensitivity of the appropriate
dual SQUID, and, thus, to estimate the suitability of the given
HTS for a high sensitivity SQUIDs building.
\section{About Scaling Behaviour of the Interferometer Sensor Noise}
Let the total interferometer  spectral noise energy density
(s.\,n.\,e.\,d.) $S_{\epsilon}$ be determined in a general way, as
usual (see, for example~\cite{Polushkin1993,Sloggett1993}){,} by
its fluctuation components $S_{\epsilon i}$, $i=1,2...,n$, which
are considered uncorrelated
\begin{equation}\label{1e}
S_{\epsilon}=S_{\epsilon 1}+ S_{\epsilon 2}+...+S_{\epsilon n}.
\end{equation}
Let, for example, $S_{\epsilon 1}$ be a s.\,n.\,e.\,d. of the
SQUID sensor, $S_{\epsilon 2}$ -- s.\,n.\,e.\,d. of the tank
circuit coupling inductively to the SQUID, $S_{\epsilon 3}$ --
s.\,n.\,e.\,d. of the feeder, $S_{\epsilon 4}$ -- s.\,n.\,e.\,d.
of the first preamplifier stage. In the best case, the components
specified differ from each other no more than one order of
magnitude~\cite{Koelle}. Let the s.\,n.\,e.\,d. of the SQUID
sensor be written as
\begin{equation}\label{2e}
S_{\epsilon 1}=S_{\epsilon 1}(L_S)=S_{\epsilon d}\cdot K(L_S),
\end{equation}
where $S_{\epsilon d}$ is the value of the SQUID s.\,n.\,e.\,d.
(energy resolution) depending on the absolute temperature and weak
link parameters, a critical current and a normal resistance, and
also, maybe, on Josephson junction capacity, but independent of
$L_S$, and $K(L_S)$ is a dimensionless scale factor depending
monotonously on $L_S$. It is supposed that all other components
$S_{\epsilon i}$, $i\neq 1$, do not depend on $L_S$. Thus,
$S_{\epsilon d}$ is s.\,n.\,e.\,d. of the SQUID sensor with a
rather small inductance. Let now $K(L_S)$ increases, together with
$L_S$, and the remaining parameters in (\ref{1e}) remain constant,
so that, when $L_S\gg L_F$, one could obtain $S_{\epsilon 1}\gg
S_{\epsilon i}$, $i=2,3,4$. Hence, from (\ref{1e}) and (\ref{2e}),
it  evidently follows:
\begin{equation}\label{3e}
S_{\epsilon}\simeq S_{\epsilon 1}(L_S)= S_{\epsilon d}\cdot
K(L_S).
\end{equation}
The last equation (\ref{3e}) reflects two important facts.  First,
in the case of high inductance of the quantization loop, the
interferometer noise is determined mainly by the SQUID sensor
noise. Second, the interferometer s.\,n.\,e.\,d. at the high
values of inductance of the quantization loop is proportional to
the s.\,n.\,e.\,d. of the SQUID sensor with the same weak link
parameters, but a small quantization loop inductance. If the
functional dependence of the $K(L_S)$ is known, in the presence of
the experimental data for $S_{\epsilon}$ of a SQUID with a large
inductance, the relationship  (\ref{3e}) allows one to determine
energy resolution $S_{\epsilon}$ of a SQUID with the same weak
link but with optimum inductance value. Let us obtain now a
relationship between the s.\,n.\,d. $S_\varphi$ of a SQUID with a
large quantization loop inductance and the s.\,n.\,d. $S_{\varphi
d}$ of a SQUID with an optimum quantization loop inductance. By
definition, the SQUID s.\,n.\,e.\,d. $S_\epsilon$ (with a
dimension of the action,
$[S_\epsilon]=\mbox{J}\cdot\mbox{Hz}^{-1}$) is connected with the
s.\,n.\,d. $S_\varphi$, $[S_\varphi]=\Phi_0/\mbox{Hz}^{1/2}$, by
the relationship
\begin{equation}\label{4e}
S_\epsilon=\frac{S^2_\varphi}{2L_S}
\end{equation}
It has been experimentally proved~\cite{Polushkin1989}{,} that the
optimum quantization loop inductance value is close to $L_F$.
Taking into account the equation
\begin{equation}\label{5e}
S_{\epsilon d}=\frac{S^2_{\varphi d}}{2L_F},
\end{equation}
and, combining (\ref{3e}), (\ref{4e}), (\ref{5e}), it is easy to
obtain the useful estimate
\begin{equation}\label{6e}
S_{\varphi d}\simeq S_{\varphi}\cdot\left[\frac{L_F}{L_S\cdot
K(L_S)}\right]^{1/2}{,}\mbox{ }L_S\gg L_F.
\end{equation}
Let us regard the (\ref{6e}) as s.\,n.\,d. of dual SQUID, i.\,e.
\begin{equation}\label{7e}
S_{\varphi\mbox{ }dual}\simeq
S_{\varphi}\cdot\left[\frac{L_F}{L_S\cdot
K(L_S)}\right]^{1/2}{,}\mbox{ }L_S\gg L_F.
\end{equation}
Assuming that $K(L_S)=\exp{[(L_S/2L_F)-1/2]}$, we obtain
\begin{equation}\label{8e}
S_{\varphi\mbox{ }dual}\simeq
S_{\varphi}\cdot\left(\frac{L_F}{L_S}\right)^{1/2}\cdot\exp{\left[\frac{1}{4}\cdot\left(1-\frac{L_S}{L_F}\right)\right]},\mbox{
}L_S\gg L_F.
\end{equation}
Thus, we have obtained the relationship (\ref {8e}) which links
$S_\varphi$ of a SQUID with a large quantization loop inductance
value of $L_S\gg L_F$, with $S_{\varphi\mbox{ }dual}$ of a dual
SQUID with the optimum quantization loop inductance value of $\sim
L_F$. This relation allows one to estimate, using the experimental
data, the suitability of a given high--$T_c$ material, from which
the large inductance SQUIDis was made, for building
high--sensitive interferometers.\\\indent In Table~\ref{Table1},
the experimental data for real RF SQUID parameters are given,
including loop quantization inductances $L_S $, coupling
coefficient $k$ measured by the
technique~\cite{Polushkin1989_JINR_Rapid}{,} and the values of
SQUID hysteresis parameter $\beta$ and s.\,n.\,d. $ S_\varphi$,
measured at pumping frequency indicated in the last column. The
critical current of contact $I_c $ is calculated then from $
\beta=2\pi L_S I_c/\Phi_0$, and the flux sensitivity $S_
{\varphi\mbox{ }dual}$ for the appropriate dual SQUID is
calculated from (\ref{8e}). The RF SQUID transfer function grows
with the growth of the pumping frequency $f$ as $f^{1/2}$, and RF
SQUID s.\,n.\,d. decreases with the same rate; it is necessary to
take into account these dependencies when comparing quality of
SQUIDs functioning at different pumping
frequencies~\cite{Kurkijarvi_1986}. An appropriate technique of
SQUID signal characteristics ranging, by the pumping frequency,
will be considered elsewhere~\cite{Sermyagin_2003}{.}
\begin{center}
\begin{table}
\caption{} \label{Table1}
\begin{tabular}{cccccccc}\hline
  \#&\begin{tabular}{c}$L_S$,\\nH\end{tabular}&$k$&$\beta$&\begin{tabular}{c} $I_c$,\\ $\mu$A\end{tabular} &\begin{tabular}{c} $S_\varphi$,\\ $\Phi_0$/Hz$^{1/2}$\end{tabular} &\begin{tabular}{c} $S_{\varphi\mbox{ }dual}$,\\ $\Phi_0$/Hz$^{1/2}$\end{tabular} &\begin{tabular}{c}$f_{L_tC_t}$,\\ MHz\end{tabular}  \\
\hline
  1& 1{.}3 & 0{.}12 & 18 & 4{.}6 &$3{.}1\cdot 10^{-3}$&$4{.}6\cdot 10^{-5}$&$\sim 20$\\
  2 & 1{.}61 & 0{.}28 &52 & 8{.}2 & $7{.}3\cdot 10^{-3}$&$4{.}6\cdot 10^{-5}$& ---"---\\
  3& 2 & 0{.}15 & 21 & 3{.}3 & -- & -- & ---"--- \\
  4& 2{.}76 & 0{.}22 &60&7{.}2&$1,5\cdot 10^{-3}$&$4{.}3\cdot 10^{-7}$&---"---\\
  5& 3{.}8 & -- & -- & -- & 0{.}1 & $1{.}9\cdot 10^{-6}$ &---"---\\
  6& 6{.}6 & 0{.}101 &7  &0{.}35  & $\sim 0{.}1$ &$\sim 1{.}7\cdot 10^{-9}$&---"---\\
  7& 24 & 0{.}075 & -- & 0{.}04 & -- & -- &---"---\\
  8 & 0{.}51 & -- & 2     & -- &$10^{-3}$   &$1{.}6\cdot 10^{-4}$  &---"---~\cite{Polushkin1989}\\
  9 & 0{.}7 & -- & -- & -- &$3\cdot10^{-4}$   &$2{.}7\cdot 10^{-5}$
&25~\cite{Il'ichev1997}\\
  10 & 0{.}45 & -- & 1{.}5 &0{.}835 &$7\cdot10^{-5}$ & $1{.}4\cdot10^{-5}$ &300~\cite{Chesca_1998} \\
  11 & 0{.}9  & -- & 2{.}25&0{.}835 &$5{.}66\cdot10^{-4}$&$2{.}7\cdot10^{-5}$&---"---\\ \hline
\end{tabular}
\end{table}
\end{center}
\section{A Comment on the Table}
For SQUIDs $\#\#$1$\div$7, the experimental data obtained by the
author of the present paper are given, other data are taken from
references.\\\indent The data are not complete; s.\,n.\,d. for
$\#$2 and $\#$7 are not retained. Parameters for $\#$7 were
defined with an error difficult to estimate, therefore the real
value of $L_S$ could differ from the specified; most likely it was
smaller. Data of $k$, $\beta$, and $I_c $ for $\#$5 are lost as
well; for SQUIDs $\#\#$8 $\div$11, the reader is kindly addressed
to the references. However, the data omitted are not crucial for
conclusions of the qualitative theory submitted. It can be seen
that in the range $L_S<2$\,nH, for $S_{\varphi\mbox{ }dual}$, real
values can be obtained corresponding to the self--noise of the
optimum SQUID made from the sertain HTS. For $\#$8,  the value of
$S_ {\varphi\mbox{ }dual}\sim 10^{-4}$ \,$\Phi_0$/Hz$^{1/2}$ is
not very high, which is usual for bulk SQUIDs made of ceramics
typical to the initial period of HTS research in the late
eighties, though already sufficient for applications. $\#$1 and
$\#$2 are created by the painting method, and by far overcome the
psychological barrier of $10^{-4}$\,$\Phi_0$/Hz$^{1/2}$, into the
value area of units $10^{-5}$\,$\Phi_0 $/Hz$^{1/2}$, close to the
best theoretical estimate~(\ref {25}), near which the thin film RF
SQUIDs $\#$9, $\#$10 and $\#$11 are situated; these latter two
were operated at the pumping frequency of 300\,MHz. Hence, sample
$\#$10 has the weak link quality close to our samples $\#$1 and
$\#$2.\\\indent
For SQUIDs whith a very large inductance
$L_S>2$\,nH, the extrapolation to the dual SQUIDs does not work
thus far due the very low level of the noise predicted. It means,
that in RF SQUIDs with a very large  quantization loop inductance
values, quantum interference is suppressed not so strongly as it
follows from the simple formula (\ref{8e}); our assumption of the
suppression character (\ref{L>L_F}) of quantum interference with
growth of inductance $L_S$, appears inexact in this area. One of
the possible ways of solving this problem consists in a consequent
account of the reduction phenomena of the RF SQUID transfer
function, with the fluctuation level growing~\cite{Il'ichev1997,
Chesca_1998, Greenberg_1999, Enpuku1994}\phantom{}; this concerns
both thermal and quantum fluctuations~\cite{Sermyagin_1997_1}{.}
The general conception is that with the reduction of SQUID
transfer function, the SQUID sensitivity decreases both to the
useful signal and to fluctuations.\\ The interference in $\#$6 and
$\#$7 was observed, despite of extremely high SQUIDs inductance
values $L_S$ and small values of critical current $I_c$. These
values were much less the fluctuation current value of 3\,$\mu$A,
which is considered the lower value of a critical current for
isolated HTS--based junction at 77\,K, admitting the observing of
the quantum interference.
\section{Conclusions}
The fluctuation inductance $L_F$ is not an upper limit of the
quantization loop inductance, exceeding which the quantum
interference becomes unobservable and SQUID ceases to operate.
More likely, one should consider $L_F$ as a quality criterion of
the SQUID, actually as an optimum value of the quantization loop
inductance. So far the large inductance SQUIDs with $L_S\gg L_F $
were considered not applicable as supersensitive magnetic flux
sensors, due to a large level of their self--noise. However, this
feature allows one to receive an estimate of the maximum
sensitivity of the RF SQUID with an optimum inductance of the
quantization loop and made of appropriate HTS. The obtained
estimate (\ref {8e}) of the $S_\varphi(L_S)$ dependence based on a
rough model (\ref{L>L_F}), appears true, if compared to the
obtained experimental data for the large inductance values of
$2\cdot10^{-9}>L_S>10^{-10}$, H. However, an interpretation of the
experimental data for higher inductance values of
$L_S>2\cdot10^{-9}$\,H needs a new approach (see e.\,g.
\cite{Sermyagin_2003}).

\section{Acknowledgments}
The results presented could not be obtained without an intensive
development of the painting on method and HTS--sample
investigations carried out by Prof.\,N.V.\,Vuong and
Dr.\,E.V.\,Pomyakushina (Raspopina), under supervision of
Prof.\,B.V.\,Vasiliev. Also, I wish to thank Dr.\,M.B.\,Miller and
O.Yu.\,Tokareva for their support and help.

\end{document}